\documentclass[aps,pra,reprint,amsmath,amssymb,superscriptaddress,nobibnotes, longbibliography]{revtex4-1}
\usepackage{graphicx,SIunits}
\usepackage{bm}     
\usepackage{hyperref} 
\usepackage{dsfont}    
\usepackage{color}

\usepackage{lineno}

\usepackage{amsmath}
\usepackage{amsthm}
\usepackage{amssymb}
\usepackage{tikz}

\usetikzlibrary{decorations.pathreplacing}
\usepackage{braket}
\usepackage{dsfont}

\usepackage{color}
\usepackage{lineno}
\expandafter\let\csname equation*\endcsname\relax
\expandafter\let\csname endequation*\endcsname\relax

\begin{document}

	\title{Experimental demonstration of optimal measurement for unambiguously discriminating asymmetric qudit states}
	\author{Kang-Min Hu}
    \email{These authors contributed equally to this work.}
	\affiliation{Center for Quantum Technology, Korea Institute of Science and Technology (KIST), Seoul 02792, Korea}
    \affiliation{Division of Quantum Information, KIST School, University of Science and Technology, Seoul 02792, Korea}
	
	\author{Min Namkung}
    \email{These authors contributed equally to this work.}
	\affiliation{Center for Quantum Technology, Korea Institute of Science and Technology (KIST), Seoul 02792, Korea}
	
	\author{Myung-Hyun Sohn}
	\affiliation{Department of Applied Physics, Kyung Hee University, Yongin 17104, Korea}
	
	\author{Hyang-Tag Lim}
	\email{hyangtag.lim@kist.re.kr}
	\affiliation{Center for Quantum Technology, Korea Institute of Science and Technology (KIST), Seoul 02792, Korea}
	\affiliation{Division of Quantum Information, KIST School, University of Science and Technology, Seoul 02792, Korea}

\date{\today} 
	
\begin{abstract}
Identification of nonorthogonal quantum states without error is crucial for various applications in quantum information technology, as well as the foundations of quantum physics. Theoretical studies have proposed measurements that maximize the success probability of unambiguously discriminating quantum states. However, these methods are not always experimentally feasible, which has led most demonstrations to focus on equiprobable symmetric states. Here, we establish a projective measurement scheme that optimally discriminates multiple asymmetric qudit states. We experimentally demonstrate this optimal projective measurement using a photonic orbital angular momentum state, where asymmetric qudit states are encoded in the Laguerre-Gaussian modes of a heralded single-photon state. Our results have broad applications in high-dimensional quantum state-based quantum information processing, including quantum key distribution and quantum sensing.
\end{abstract}
	
	
\maketitle
\section{Introduction}
States of a quantum system are generally nonorthogonal, making it impossible for measurements to perfectly determine which state has been prepared. Instead, one can consider a measurement that identifies the prepared state with the maximum success probability allowed by quantum theory. From this perspective, quantum state discrimination is highly relevant to quantum foundations \cite{j.a.bergou,j.a.bergou_text,s.m.barnett}. Useful strategies include minimum-error state discrimination (MESD) \cite{c.w.helstrom,s.m.barnett_med,j.bae,d.ha_med,d.ha_med2}, unambiguous state discrimination (USD) \cite{i.d.ivanovic,d.dieks,a.peres,g.jaeger,y.c.eldar,m.a.jafarizadeh,a.chefles,a.chefles2,s.pang,h.sugimoto,j.a.bergou3,d.ha}, maximal-confidence discrimination \cite{s.croke,u.herzog,u.herzog2,o.jiminez_mcd,h.lee_mcd}, error margin discrimination \cite{m.a.p.touzel_em,a.hayashi_em,h.sugimoto_em}, and discrimination with a fixed rate of inconclusive outcomes \cite{a.chefles_frir,c.-w.zhang_frir,j.fiurasek_frir}. Notably, USD holds great promise not only for rigorously interpreting quantum nature~\cite{m.n.bera} but also for advancing quantum information tasks, such as quantum communication~\cite{c.h.bennett,k.banaszek,f.e.becerra}, quantum random number generation \cite{j.b.brask,h.tebyanian2}, and quantum sensing \cite{r.lecamwasam}.

The purpose of USD is to establish an unambiguous measurement that identifies quantum states without error. For geometrically symmetric states, it is possible to analytically determine an unambiguous measurement that maximizes the success probability of state discrimination~\cite{a.chefles2}. However, quantum states are generally asymmetric. Therefore, the structure of unambiguous measurements must be extended to accommodate asymmetric states. Several works have theoretically proposed unambiguous measurements that optimally discriminate three asymmetric qutrit states~\cite{h.sugimoto,s.pang,j.a.bergou3,d.ha}. However, measurements implemented in experiments are often constrained by the limited structure of positive operator-valued measures (POVMs). In fact, experiments reported in several studies have primarily focused on discriminating a few symmetric states~\cite{r.b.m.clarke,o.jiminez,m.a.solisprosser,j.s.sidhu,l.f.melo}. This highlights the need to develop a measurement that performs optimal USD for asymmetric quantum states.

Several works have proposed that POVMs for the USD are established by enlarging the state space and performing a projective measurement on the enlarged space~\cite{p.-x.chen,y.sun,y.sun2,m.mohseni}. Notably, a photonic system using an orbital angular momentum (OAM) state can proficiently implement projective measurements to discriminate symmetric states~\cite{s.franke-arnold2,m.agnew,s.goel}, as it efficiently supports a high-dimensional space. This advantage suggests the possibility that \textit{the discussed scheme may also be applicable for optimally discriminating arbitrary asymmetric qudit states.} 

Stressing the arbitrariness of qudit states and their prior probabilities is particularly important in practical quantum communication~\cite{g.cariolaro}. The proportions of dits constituting an entire message are typically non-uniform. It implies that, during the quantum communication, the prior probabilities of the encoded qudit states are not equal to each other. This perspective also affects the amount of information shared between a sender and a receiver, relevant to secret key rate of USD-based quantum communication~\cite{m.namkung_q}. Moreover, unequal prior probabilities of quantum states in a given ensemble can influence the nonclassicality of the described quantum system~\cite{d.ha_nonl,j.shin_cont}, demonstrating interesting features of the quantum foundation~\cite{c.h.bennett_nonl,d.schmid_cont,k.flatt_cont}.

In this work, we demonstrate the structure of projective measurements that are useful for \textit{optimally discriminating multiple asymmetric qudit states}, experimentally implemented using an OAM state of a single-photon. Specifically, we show that a four-dimensional projective measurement can achieve optimal unambiguous discrimination among three asymmetric qutrit states. Beyond qutrit states, we extend the application of projective measurements to optimal USD for multiple asymmetric qudit states. It suggests that a generalized measurement for discriminating high-dimensional asymmetric qudit states is efficiently established using the projective measurement, thereby circumventing the aforementioned infeasible interaction. In our experimental realization, three qutrit states, encoded with Laguerre-Gaussian (LG) modes, are discriminated using a four-dimensional projective measurement implemented with a spatial light modulator (SLM). Our experiment successfully achieves the maximum success probability for USD, as predicted by theoretical evaluations.

\begin{figure}
\centerline{\includegraphics[width=0.8\columnwidth]{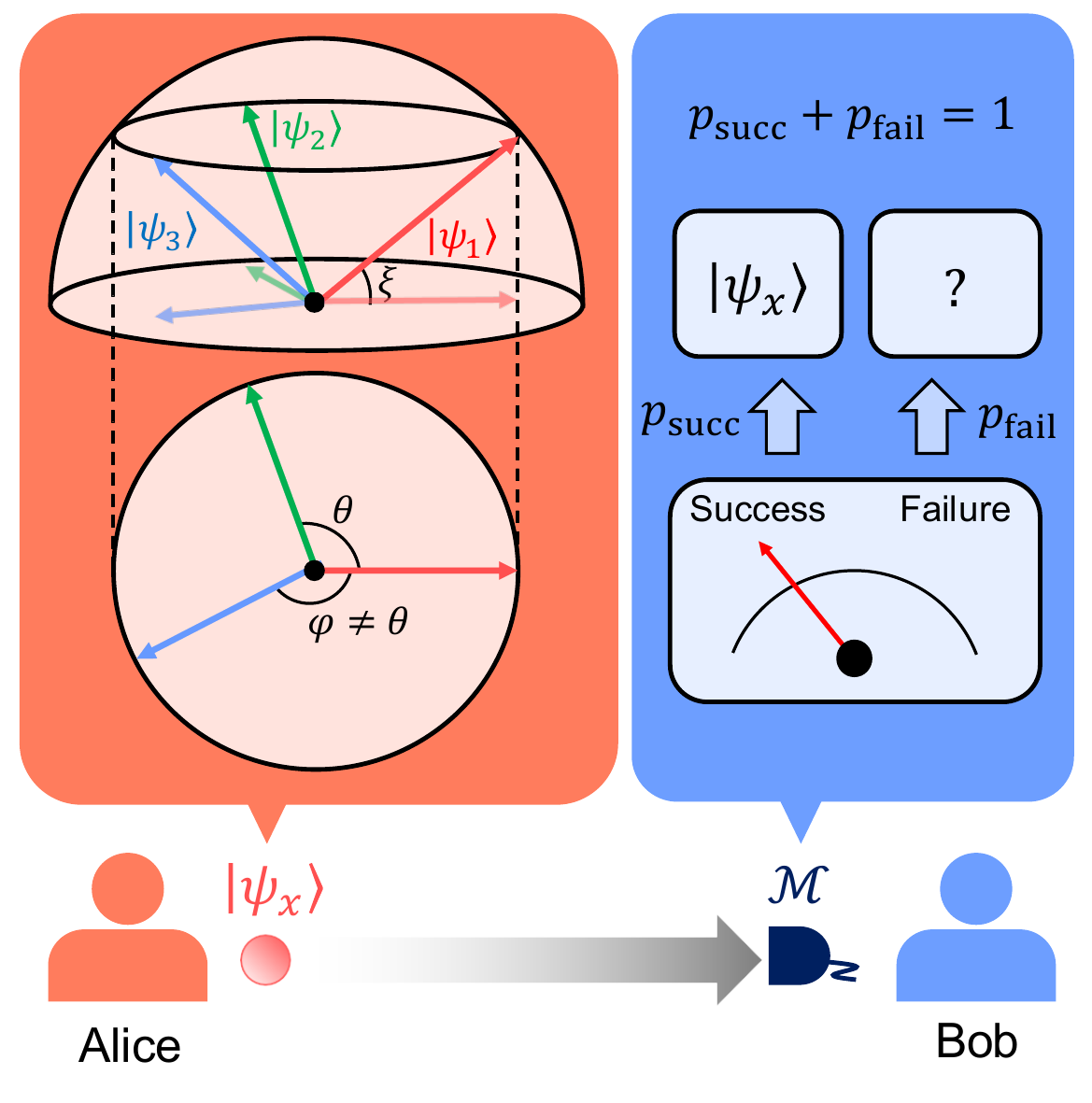}}
\caption{Concept of USD, where Alice prepares one of three qutrit states. The two angles $\varphi$ and $\theta$ in Alice's quantum states are distinct, indicating asymmetry.  Alice prepares a quantum state $|\psi_x\rangle$, and Bob performs a measurement $\mathcal{M}$ to identify the prepared state. There are only two possible outcomes in USD. If the measurement outcome is conclusive ($y\not=?$) with a probability $p_{\rm succ}$, Bob can confidently determine that Alice prepared $|\psi_x\rangle$. Otherwise, with probability $p_{\rm fail}=1-p_{\rm succ}$, the measurement outcome provides no useful information.}
\centering
\label{fig:1}
\end{figure}

\section{Theory}

We begin by introducing the quantum state discrimination scenario. Consider that a quantum state $|\psi_x\rangle$ is prepared from the set $\{|\psi_1\rangle,\cdots,|\psi_d\rangle\}$ in a state space, with a prior probability $q_x$. To identify which state is prepared, a measurement represented by a POVM $\{\hat{M}_y\}_{y\in\{1,\cdots,d\}\cup\{?\}}$ is performed. Here, $\hat{M}_y$ is the element corresponding to the outcome $y$. If $y\in\{1,\cdots,d\}$ is obtained with probability $\langle\psi_y|\hat{M}_y|\psi_y\rangle$, the measurement identifies the prepared state as $|\psi_y\rangle$. Otherwise, for an inconclusive outcome $y=?$, it is impossible to extract any information about the prepared state.  In this scenario, the success probability of identifying the prepared state is given by $P_s=\sum_{y=1}^{d}q_y\langle\psi_y|\hat{M}_y|\psi_y\rangle$. The main goal of quantum state discrimination is to find an optimal POVM that maximizes the success probability.

In USD, an unambiguous measurement, described by a POVM $\big\{\hat{M}_y^{\rm (USD)}\big\}_{y\in\{1,\cdots,d\}\cup\{?\}}$ on the state space, is established such that $\langle\psi_x|\hat{M}_{y\not=x}^{\rm (USD)}|\psi_x\rangle=0$. This condition implies that all conclusive outcomes $(y\not=?)$ indicate $|\psi_x\rangle$ without error (see Fig.~\ref{fig:1}). To satisfy this constraint, the operator $\hat{M}_y^{\rm (USD)}$, corresponding to a conclusive outcome, takes the form  
$\hat{M}_y^{\rm (USD)}=\alpha_y|\widetilde{\psi}_y\rangle\langle\widetilde{\psi}_y|$, where $\alpha_y\in[0,1]$ and $\langle\widetilde{\psi}_y|\psi_x\rangle=\delta_{xy}$ for all $x,y\in\{1,\cdots,d\}$ \cite{s.pang,h.sugimoto,j.a.bergou3,d.ha}. Subsequently, $\hat{M}_{?}^{\rm (USD)}$, corresponding to an inconclusive outcome, is given by $\hat{M}_{?}^{\rm (USD)}=\hat{I}-\sum_{y=1}^{d}\hat{M}_y^{\rm (USD)}$ due to completeness condition. The requirement $\hat{M}_?^{\rm (USD)}\ge 0$ implies the inequality $\mathrm{G}_d-\mathrm{diag}\left(\alpha_1,\cdots,\alpha_d\right)\ge 0$, where $\mathrm{G}_d=\left\{\langle\psi_x|\psi_{x'}\rangle\right\}_{x,x'=1}^{d}$ is the Gram matrix~\cite{j.a.bergou,s.pang,d.ha}. In other words, the unambiguous measurement corresponds to a real vector on a closed convex set: 
\begin{eqnarray*}
    \mathcal{C}_d=\left\{\left(\alpha_1,\cdots,\alpha_d\right):\mathrm{G}_d-\mathrm{diag}\left(\alpha_1,\cdots,\alpha_d\right)\ge 0\right\}.
\end{eqnarray*}
Due to this convexity, an optimal measurement can be represented as the tangential point between the convex set and the hyperplane of success probability $P_s=\sum_{y=1}^{d}q_y\alpha_y$.

The realization of the optimal POVM generally requires an interaction between the state to be measured and an ancillary state~\cite{m.a.neumark}. However, such interactions can be challenging to implement in optical systems~\cite{p.kok}, limiting the applicability of these measurements in various quantum technologies. Here, we propose methodology to implement the POVM by performing a projective measurement on the extended state space, described by an orthonormal basis $\{|D_y\rangle\}_{y\in\{1,\cdots,d\}\cup\{?\}}$, to circumvent the need for complicated interactions. A basis vector $|D_y\rangle$ is proportional to $|\widetilde{\psi}_y\rangle+|\phi_y\rangle$, where $|\widetilde{\psi}_y\rangle$ is derived from the element $\hat{M}_y^{\rm (USD)}$, and $|\phi_y\rangle$ is orthogonal to $|\widetilde{\psi}_y\rangle$. Consequently, it is verified that $|\langle D_y|\psi_x\rangle|^2=0$ for all $x\not=y$, indicating that USD is achievable {(The detailed information is contained in {Appendix A})}. 

We optimize the vectors $|\phi_y\rangle$ in $|D_y\rangle$ under the constraint $\langle D_y|D_{y'}\rangle=\delta_{yy'}$, aiming to maximize the success probability $P_s=\sum_{y}q_y|\langle D_y|\psi_y\rangle|^2$. We note that a set of projective measurements does not have convex property. It implies that the convex optimization techniques frequently used in quantum state discrimination \cite{j.bae,s.boyd} are not applied to optimize the measurement basis. We first consider the USD of two qubit states $|\psi_1\rangle$ and $|\psi_2\rangle$. To discriminate these two states, a three-dimensional projective measurement, consisting of an orthonormal basis $\{|D_1\rangle,|D_2\rangle,|D_?\rangle\}$, can generally be employed. Here, a basis vector $|D_y\rangle$ for $y\in\{1,2\}$ is proportional to $|\widetilde{\psi}_y\rangle+a_y|\phi_y\rangle$, where $|\phi_y\rangle$ is a normalized vector orthogonal to $|\widetilde{\psi}_y\rangle$. The optimal projective measurement is determined by solving the following optimization problem:
\begin{eqnarray*}
    \mathrm{maximize} && \ P_s=q_1|\langle D_1|\psi_1\rangle|^2+q_2|\langle D_2|\psi_2\rangle|^2,\nonumber\\
    \mathrm{subject \ to} && ~\langle \widetilde{\psi}_1|\widetilde{\psi}_2\rangle+a_1a_2=0. 
\end{eqnarray*}
The analytical solution provides the maximum success probability, commonly referred to as the Ivanovic-Diek-Peres limit \cite{i.d.ivanovic,d.dieks,a.peres,g.jaeger}. If $\sqrt{q_2/q_1}$ is greater than $|\langle\psi_1|\psi_2\rangle|$, both $|D_1\rangle$ and $|D_2\rangle$ {are formulated as
\begin{eqnarray*}
    |D_1\rangle&=&\sqrt{\frac{1-s^2}{1+\mu(1-s^2)}}\left(|\widetilde{\psi}_1\rangle+\mu|l_2\rangle\right),\nonumber\\
    |D_2\rangle&=&\sqrt{\frac{1-s^2}{1+\nu(1-s^2)}}\left(|\widetilde{\psi}_2\rangle+\nu|l_2\rangle\right),\nonumber
\end{eqnarray*}
where $\mu$ and $\nu$ are defined as
\begin{eqnarray*}
    \mu&=&\left\{\sqrt{\frac{s^2q_1q_2}{(q_1-s^2q_2)^2}}-\frac{s^2(q_1-q_2)}{(1-s^2)(q_1-s^2q_2)}\right\}^{\frac{1}{2}},\nonumber\\
    \nu&=&\frac{s}{(1-s^2)\mu},
\end{eqnarray*}
respectively. This measurement allows} discrimination between the two qubit states. Otherwise, one of the two basis, $\{|D_1\rangle\propto|\widetilde{\psi}_1\rangle,|D_?\rangle=|\psi_2\rangle\}$ and $\{|D_2\rangle\propto|\widetilde{\psi}_2\rangle,|D_?\rangle=|\psi_1\rangle\}$, establishes the optimal projective measurement. 

\begin{figure}    
\centering
\includegraphics[width=\columnwidth]{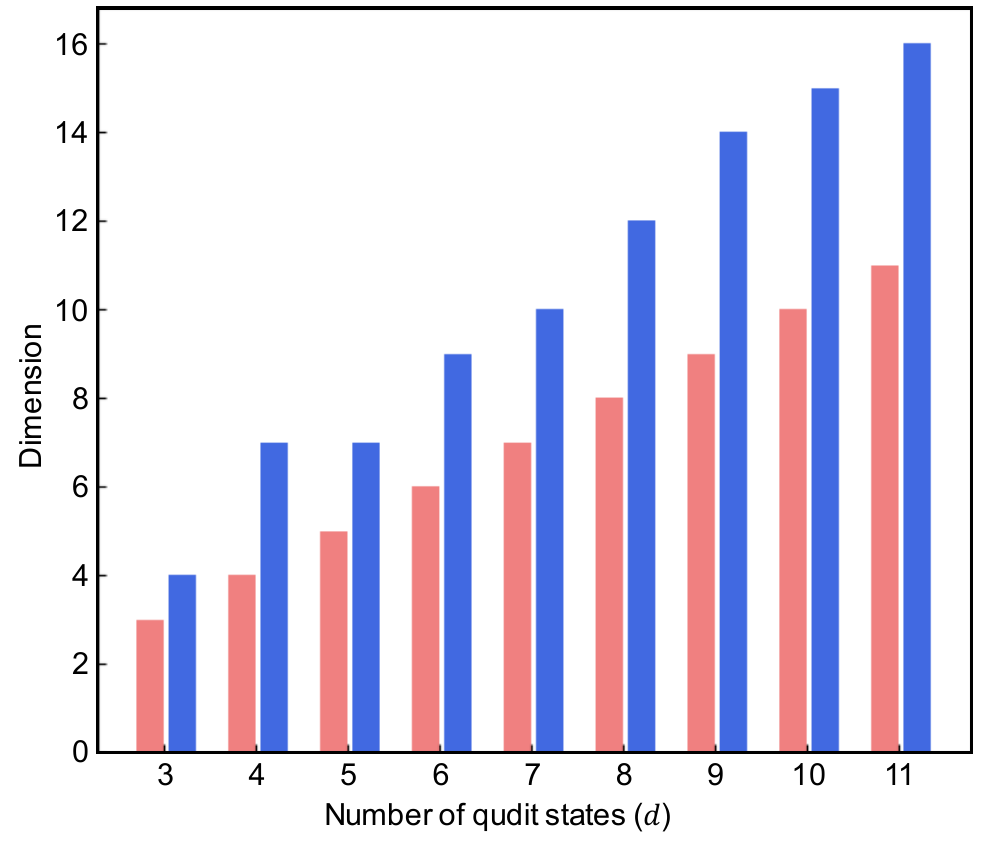}
\caption{Required dimension of a projective measurement for establishing a generalized measurement that optimally discriminates $d$ high-dimensional qudit states. Here, both red and blue bars show the dimension of a state space and that of the projective measurement, respectively.}
    \label{fig:2}
\end{figure}

\begin{figure*}
\centering
\includegraphics[width=2\columnwidth]{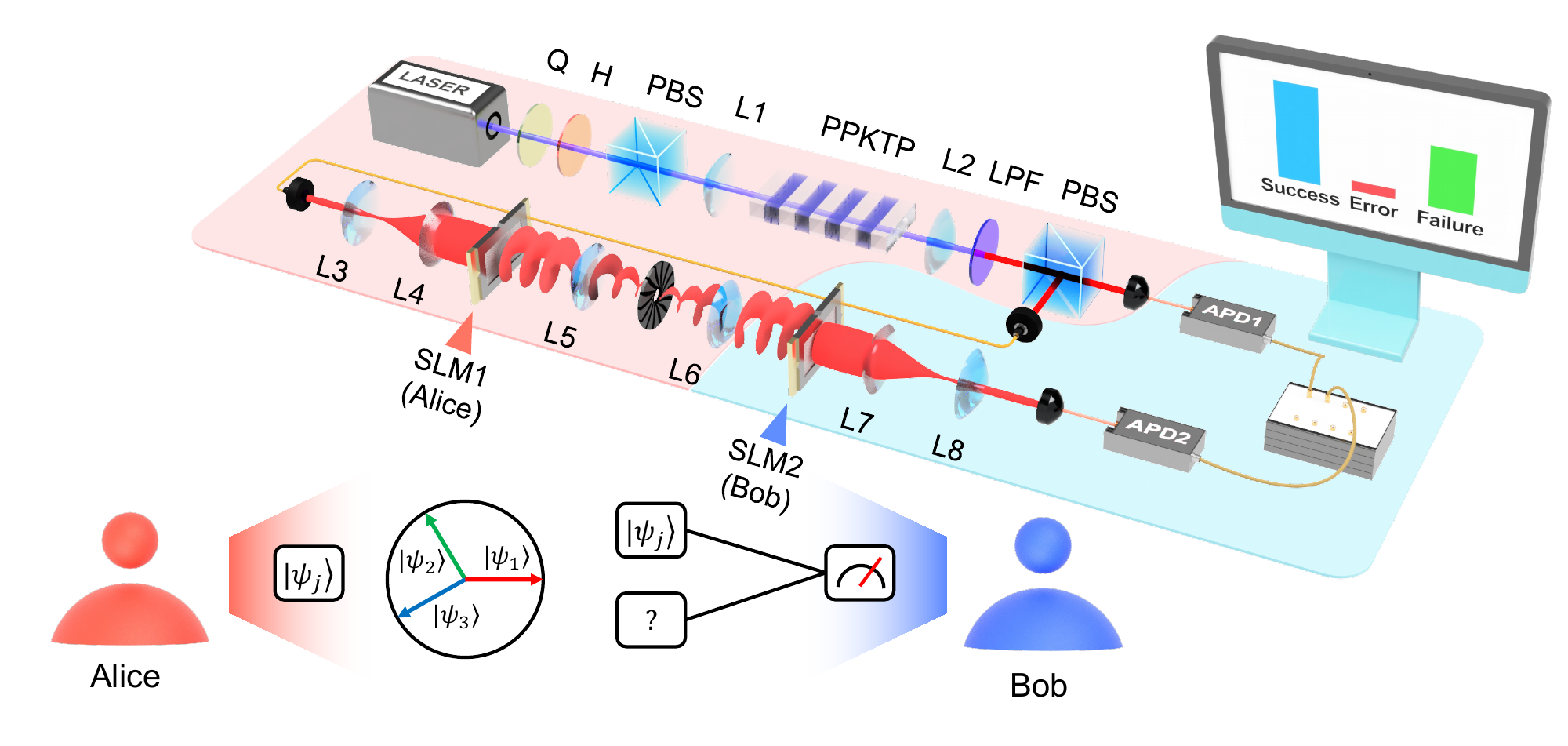}
\caption{Experimental setup for unambiguously discriminating three asymmetric qutrit states. A PPKTP crystal is pumped by a 405 nm pump laser, generating a pair of single photons via the via the type-II spontaneous parametric down-conversion process (SPDC). The idler photon is sent to APD1 for heralding, while the signal photon is directed to APD2 through an single-mode optical fiber. SLM1 is used to generate an asymmetric photon state $|\psi_j\rangle$, and SLM2 is used to discriminate the prepared states (H, Half-wave plate; Q, Quarter-wave plate;  P, Polarizing beam-splitter; L, Lens; LPF, Long pass filter; SLM, Spatial light modulator; APD, Avalanche photodiode; TCSPC, Time-correlated single-photon counter).}
    \label{fig:3}
\end{figure*}

\begin{figure*}
\centerline{\includegraphics[width=2\columnwidth]{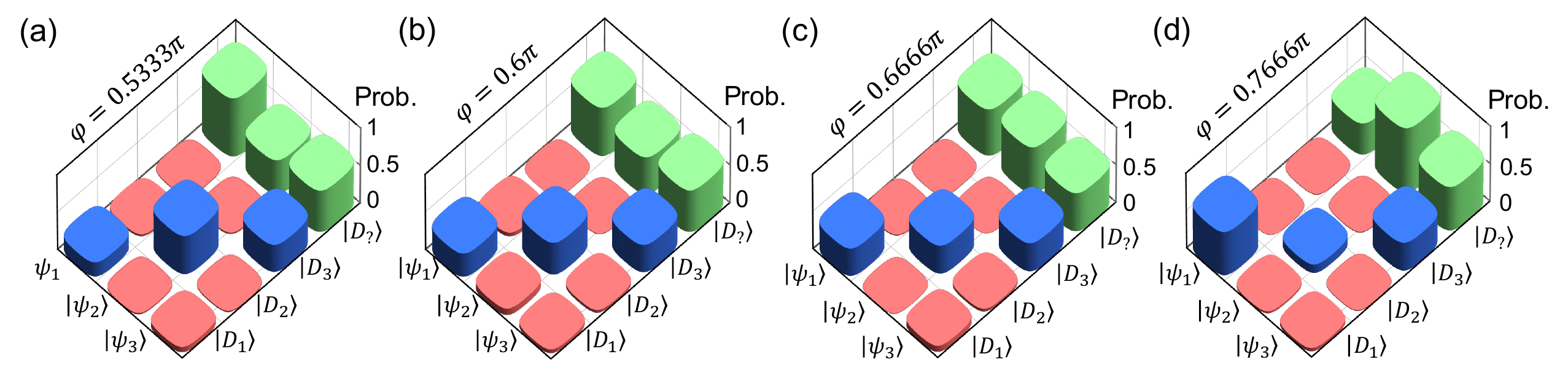}}
\caption{Experimentally obtained probability distribution for $\theta = \frac{2\pi}{3}, \xi = \frac{\pi}{3}$ and each $\varphi$. (a) $\varphi=0.53\pi$, (b) $\varphi=0.6\pi$, (c) $\varphi=0.66\pi$ and (d) $\varphi=0.76\pi$. Each measurement probability $|\langle D_k|\psi_j\rangle|^2$ is colored with blue, red, and green bars corresponding to success, error, and failure cases, respectively.}
\centering
\label{fig:4}
\end{figure*}

Moreover, the optimal projective measurements in our approach can perform optimal USD among multiple asymmetric qudit states. Throughout this article, we define qudit states to be asymmetric if at least one of the pairwise overlaps $\langle\psi_j|\psi_k \rangle$ is different from the others \cite{note}. We first verify the optimality in discriminating three asymmetric qutrit states $\{|\psi_1\rangle,|\psi_2\rangle,|\psi_3\rangle\}$, where  $\langle\psi_j|\psi_k\rangle=s_l|\epsilon_{jkl}|e^{i\zeta_l}$ with $s_l\in[0,1]$, a real number $\zeta_l$, and Levi-Civita symbol $\epsilon_{jkl}$, in general~\cite{j.a.bergou3,s.pang,d.ha}. We demonstrate that four-dimensional projective measurements, analytically described in {Appendix B}, can unambiguously discriminate these three asymmetric qutrit states with the success probability \cite{j.a.bergou3}
\begin{eqnarray*}
   P_s=q_1\frac{s_2s_3}{s_1}+q_2\frac{s_1s_3}{s_2}+q_3\frac{s_1s_2}{s_3}.
\end{eqnarray*}
Remarkably, these projective measurements also realize the optimal POVM.

Beyond the case of three qutrit states, our projective measurements enable optimal USD among multiple asymmetric qudit states. Notably, the behavior of optimal projective measurements depends on their dimensionality. When the number of states is $d=3$, a four-dimensional space suffices to establish an optimal projective measurement. Conversely, when $d>3$, the dimension of the optimal projective measurement tends to increase with $d$, unless \textit{all the states are equiprobable and symmetric.} This can be observed from an example in which all $d$ qudit states have overlaps $\langle\psi_j|\psi_k\rangle=0.3$ for $j+k=d+1$, $j=d\wedge k<d$, or $j<d\wedge k=d$, and 0.1 otherwise. Fig.~\ref{fig:2} suggests that the dimension of the projective measurement increases with $d$, and generally larger than the dimension of the state space {as detailed in Appendix B.2}. This aspect is interpreted as follows. For given $d$ states, $d+1$ measurement basis vectors are needed to establish a projective measurement to unambiguously discriminate them. Inner products among these vectors should be zero, and thus the dimension of the enlarged space is generally $\frac{d(d+1)}{2}$, leading to the additional dimension $\frac{d(d-1)}{2}$.

\section{Experiment}

Our experimental setup for unambiguously discriminating three asymmetric qutrit states is shown in Fig.~\ref{fig:3}. A heralded single-photon state is prepared via the type-II spontaneous parametric down conversion (SPDC) process using a 405 nm continuous-wave pump laser and 10-mm-long periodically-poled potassium titanyl phosphate (PPKTP) crystal. One of three qutrit states is prepared and encoded with LG modes, performed by controlling holographic images of SLM 1~\cite{bolduc}. The encoded qutrit state propagates through a lens-iris-lens (4-f) system~\cite{bent}. During propagation, the central iris allows only the first-order diffraction to pass, filtering out all other diffraction orders.

 A projective measurement is performed on the propagated state using a two-dimensional Fourier transform on the holographic image with SLM 2~\cite{rambach}. Measurement outcomes are subsequently obtained as coincidence counts detected with a single-mode optical fiber.  This process is repeated for each qutrit state to experimentally evaluate the measurement probabilities. Finally, the success probability is calculated as the sum of the evaluated probabilities, each weighted by its prior probability. For experimental details, see {Appendix C}.

We consider the following three asymmetric qutrit states: 
\begin{eqnarray*}
    |\psi_1\rangle&=&\cos\xi|l_0\rangle+\sin\xi|l_2\rangle,\\
    |\psi_2\rangle&=&\cos\xi(\cos\theta|l_0\rangle+\sin\theta|l_1\rangle)+\sin\xi|l_2\rangle,\\
    |\psi_3\rangle&=&\cos\xi(\cos\varphi|l_0\rangle-\sin\varphi|l_1\rangle)+\sin\xi|l_2\rangle,
\end{eqnarray*}
all of which are spanned by basis vectors $|l_0\rangle$, $|l_1\rangle$, and $|l_2\rangle$. Here, $l_0$, $l_1$, and $l_2$ correspond to the azimuthal indices $l \in \{-3,-1,1\}$, respectively. These qutrit states are neither on the equator nor the pole of the unit sphere in the three-dimensional real space spanned by $\{|l_0\rangle,|l_1\rangle,|l_2\rangle\}$. We set $\theta=\frac{2\pi}{3}$ and $\varphi$ such that $0<\varphi<\pi$ and $\varphi\not=\frac{2\pi}{3}$ to ensure asymmetry {(We particularly refer to the case of $\varphi=\frac{2\pi}{3}$ as symmetric, as the associated states are equidistant with each other~\cite{m.agnew,o.jimenez_equid,l.-l.li})}. This implies that the three qutrit states are linearly independent, enabling USD among them~\cite{a.chefles}.

We experimentally perform a projective measurement composed of basis vectors $|D_y\rangle\propto|\widetilde{\psi}_y\rangle+a_y|l_3\rangle$. Here, $|l_3\rangle$ is a basis vector corresponding to $l=3$, and all $|\widetilde{\psi}_y\rangle$ are spanned by $|l_0\rangle$, $|l_1\rangle$, and $|l_2\rangle$ (The detailed explicit form is provided in {Appendix B.1}. Figure~\ref{fig:4} illustrates measurement probabilities $|\langle D_y|\psi_x\rangle|^2$ with respect to specific $\varphi$. See {Appendix C.3} for additional experimental data for various $\varphi$. We observe that the error probabilities $|\langle D_{y\not=x}|\psi_x\rangle|^2$ are much smaller than the success probabilities $|\langle D_{x}|\psi_x\rangle|^2$ and are approximated zero. Consequently, our experimental results successfully demonstrate USD.

\begin{figure}    
\centering
\includegraphics[width=\columnwidth]{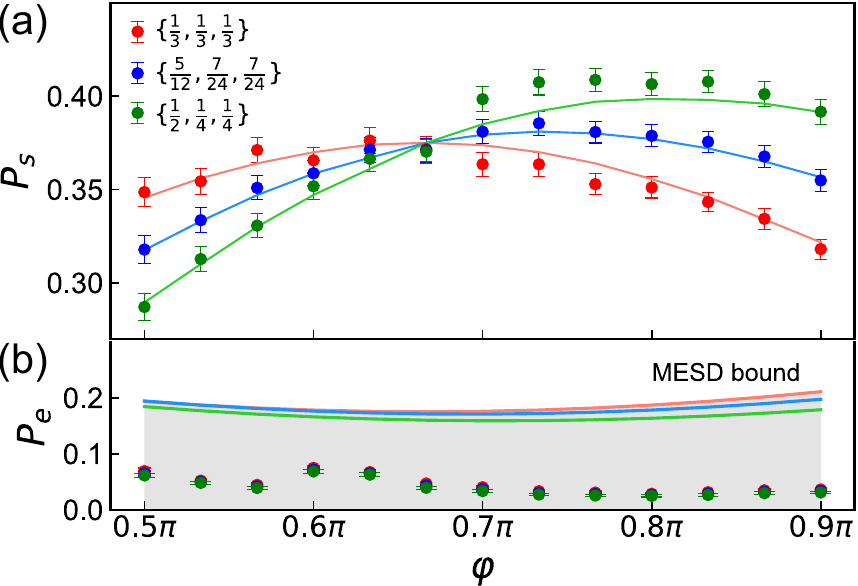}
\caption{Experimental results on success probability considering different prior probabilities for each $\varphi$. Note that the prepared states are symmetric only when  $\varphi = \frac{2}{3}\pi$. Red, blue, and green dots (lines) indicate experimental (theoretical) values for the cases of $\{q_1,q_2,q_3\}=\{\frac{1}{3},\frac{1}{3},\frac{1}{3}\}$, $\{\frac{5}{12},\frac{7}{24},\frac{7}{24}\}$, and $\{\frac{1}{2}, \frac{1}{4}, \frac{1}{4}\}$, respectively. (a) The success probability calculated from the experimental data. (b) The error probability  from the experimental data, compared with the MESD  bound. The error bars in (a) and (b) represent the standard deviation $\sigma=1$, obtained from 1000 repetitions of the Monte Carlo simulation with Poissonian errors.}
    \label{fig:5}
\end{figure}

We experimentally demonstrate the maximum success probability of discriminating all three qutrit states in Fig.~\ref{fig:5}. In Fig.~\ref{fig:5}, $\varphi$ is assumed to $0.5\pi\le\varphi\le0.9\pi$ to clearly highlight tendency around intersection of the curved lines. We observe that all data points are close to the theoretical values, indicating that our scheme effectively performs optimal USD. It is worth noting that even slight misalignments can disrupt the orthogonality of superposed LG modes, leading to a nonzero probability of error. For detailed discussion for the orthogonality, see {Appendix C.2}. To assess the performance of our experiment under this limitation, we compare the experimentally obtained error probability with the error probability in MESD \cite{c.w.helstrom,j.bae,d.ha_med}, referred to as the MESD bound. Our results show that the experimental error probability is much lower than the MESD bound, justifying that our experimental results accurately demonstrate optimal USD for the perspective of a criterion discussed in the previous studies~\cite{m.agnew,s.goel}. Moreover, the experimental error rates in Fig.~\ref{fig:5} have average of 4.24\% and do not exceed 7.46\%, which are significantly lower than the experimentally obtained success rates. It emphasizes that our experiment nearly realizes the USD even under inevitable imperfection.

Interestingly, when prior probabilities are unequal, there are the asymmetry parameters $\theta$ and $\varphi$ that can enhance the success probability of discriminating the qutrit states. For example, if $\varphi$ is greater than $\frac{2\pi}{3}$, the success probabilities for the case of $\{\frac{5}{12},\frac{7}{24},\frac{7}{24}\}$ and $\{\frac{1}{2},\frac{1}{4},\frac{1}{4}\}$ are higher than that for the uniform case $\{\frac{1}{3},\frac{1}{3},\frac{1}{3}\}$.

\section{Conclusion}
From a fundamental perspective, we provide a theoretical methodology to achieve optimal USD among multiple asymmetric quantum states. Our approach introduces a projective measurement that attains the maximum success probability allowed by quantum theory. When one of three asymmetric states is prepared, four-dimensional projective measurements effectively realize the proposed optimal USD. Moreover, the projective measurements in our methodology can achieve the maximum success probability for discriminating multiple asymmetric qudit states. We experimentally demonstrate our methodology using an OAM state of a single-photon, where asymmetric qutrit states are encoded with LG modes. Our experimental results confirm the successful realization of optimal USD. Notably, we suggest that when prior probabilities are unequal, there exists an asymmetric configuration of the qutrit states that enhance the success probability.

We emphasize that the photonic OAM platform used in our work is promising for implementing qudit-based protocols in quantum information processing~\cite{j.c.garcia,y.chen,z.qi,z.sun,b.-j.kim,z.-q.zhang,x.-h.zhan}. We have focused on unambiguous discrimination of pure qudit states {encompassing previous works~\cite{r.b.m.clarke,j.b.brask,m.agnew,s.gomez,w.cai,g.a.steudle}}, with a future extension to mixed states whose support spaces are not fully overlapping, partially inspired from the ideas in Refs.~\cite{t.rudolph,u.herzog3,u.herzog4}. Moreover, our scheme is employable for discriminating arbitrary multiple qudit states with feasibly controlling a measurement, enhancing the practicality of the discrimination task. Therefore, we believe that our methodology and results are valuable for advancing quantum technologies, beyond calculating the optimal POVM of the state discrimination. In this promising context, further experimental demonstration to simultaneously discriminate high-dimensional asymmetric qudit states represents a meaningful next step. Additionally, as successfully demonstrated by Ref.~\cite{s.goel}, it is intriguing to explore simultaneous discrimination of all asymmetric qudit states using the OAM platform.

\begin{widetext}
    \section*{Appendix A. Detailed background}
    \subsection*{A.1. General structure}
We first review a geometric structure of the measurement for unambiguously discriminating $d$ linearly independent quantum states $|\psi_1\rangle,\cdots,|\psi_d\rangle$, which was proposed in Refs. \cite{j.a.bergou,s.pang,d.ha,m.namkung}. Let us consider $d+1$ positive operator-valued measure (POVM) elements
\begin{equation}\label{povm}
    \hat{M}_j=\alpha_j|\widetilde{\psi}_j\rangle\langle\widetilde{\psi}_j|, \ \ \hat{M}_?=\hat{I}-\sum_{j=1}^{d}\hat{M}_{j},
\end{equation}
where $|\widetilde{\psi}_j\rangle$ is a vector described as
\begin{equation}\label{psitilde}
    |\widetilde{\psi}_j\rangle=\sum_{k=1}^{d}\mathrm{G}_{jk}^{-1}|\psi_{k}\rangle,
\end{equation}
with $\alpha_j\ge 0$ and a Gram matrix $\mathrm{G}=[\langle\psi_j|\psi_k\rangle]_{j,k=1}^{d}$. It was shown that $\hat{M}_?\ge 0$ is equivalent to $[\langle\psi_j|\hat{M}_?|\psi_k\rangle]_{j,k=1}^{d}\ge 0$ \cite{j.a.bergou}. This proposition leads us to conclude that an unambiguous measurement in Eq.~(\ref{povm}) is represented as a real vector $(\alpha_1,\cdots,\alpha_d)$ in a closed convex set:
\begin{equation}\label{C}
    \mathcal{C}_d:=\{(\alpha_1,\cdots,\alpha_d):f_d(\alpha_1,\cdots,\alpha_d):=\mathrm{det}\left[\mathrm{G}-\mathrm{diag}(\alpha_1,\cdots,\alpha_d)\right]\ge0\}.
\end{equation}

An unambiguous measurement that maximizes the success probability is described by the solution to the following convex optimization problem:
\begin{align}\label{opt}
    \mathrm{maximize} \  &  \ P_s(\alpha_1,\cdots,\alpha_d)=q_1\alpha_1+q_2\alpha_2+\cdots+q_d\alpha_d,\nonumber\\
    \mathrm{subject \ to} \ & \ (\alpha_1,\cdots,\alpha_d)\in\mathcal{C}_d.
\end{align}
Here, $q_j$ is a prior probability that $|\psi_j\rangle$ is prepared. An optimal point $(\alpha_1^{\star},\cdots,\alpha_d^{\star})$ is called \textit{the surface solution} if all $\alpha_j^{\star}$ are positive-definite. If the surface solution is optimum, then an optimal strategy is to discriminate all the quantum states. For large $d$, it is challenging to analytically solve the optimization problem in Eq.~(\ref{opt}) in general. Nevertheless, due to the convexity of the problem, it is guaranteed that a numerical method can find a global optimum.

\subsection*{A.2. Optimal measurement for discriminating two states}

When one of two quantum states $|\psi_1\rangle,|\psi_2\rangle$ is prepared, the surface solution is proven to achieve the Ivanovic-Dieks-Peres (IDP) limit \cite{i.d.ivanovic,d.dieks,a.peres,g.jaeger}. Here, the convex set $\mathcal{C}_2$ is composed of real vectors $(\alpha_1,\alpha_2)$ such that
\begin{eqnarray}
    f_2(\alpha_1,\alpha_2)=(1-\alpha_1)(1-\alpha_2)- s^2\ge 0, \ 0\le\alpha_1,\alpha_2\le1-s^2,
\end{eqnarray}
with $s=|\langle\psi_1|\psi_2\rangle|$. An optimal solution, which is a tangential point between $\mathcal{C}_2$ and the success probability, satisfies
\begin{eqnarray}
    \frac{\partial P_s(\alpha_1,\alpha_2)/\partial \alpha_1}{\partial P_s(\alpha_1,\alpha_2)/\partial \alpha_2}=\frac{\partial f_2(\alpha_1,\alpha_2)/\partial\alpha_1}{\partial f_2(\alpha_1,\alpha_2)/\partial\alpha_2}, \ f_2(\alpha_1,\alpha_2)=0.
\end{eqnarray}
By solving these equations, the optimal solution is analytically evaluated as \cite{m.namkung}
\begin{eqnarray}
    (\alpha_1^{\star},\alpha_2^{\star})=\begin{cases}
        \left(1-\sqrt{\frac{q_2}{q_1}}s,1-\sqrt{\frac{q_1}{q_2}}s\right), \ s\le\sqrt{\frac{q_2}{q_1}}\le\frac{1}{s}, \\
        \left(1-s^2,0\right) \ \mathrm{or} \ \left(0,1-s^2\right), \ \mathrm{otherwise},
    \end{cases}
\end{eqnarray}
which exhibits the optimal success probability equal to the IDP limit. Particularly, $(\alpha_1^{\star},\alpha_2^{\star})$ is a surface solution when $q_1$ and $q_2$ satisfy $s\le\sqrt{\frac{q_2}{q_1}}\le\frac{1}{s}$.

\subsection*{A.3. Optimal measurement for discriminating three states}

In the case of discriminating three quantum states $|\psi_1\rangle,|\psi_2\rangle,|\psi_3\rangle$, where $\langle\psi_j|\psi_k\rangle=s_le^{i\epsilon_{jkl}\varphi_l}$ with the Levi-Civita symbol $\epsilon_{jkl}$, all points $(\alpha_1,\alpha_2,\alpha_3)\in\mathcal{C}_3$ satisfy \cite{j.a.bergou,m.namkung,m.namkung2}
\begin{align}
    &f_3(\alpha_1,\alpha_2,\alpha_3)=(1-\alpha_1)(1-\alpha_2)(1-\alpha_3)-s_1^2(1-\alpha_1)-s_2^2(1-\alpha_2)-s_3^2(1-\alpha_3)+2s_1s_2s_3\cos(\Phi)\ge 0,\nonumber\\
    &0\le\alpha_l\le 1, \ \ (1-\alpha_j)(1-\alpha_k)\ge s_l^2, \ \ \forall j\not=k\not=l.
\end{align}
Here, $\Phi=\varphi_1+\varphi_2+\varphi_3$ is the Berry phase \cite{j.a.bergou,m.v.berry}. The above inequalities constitute a convex set $\mathcal{C}_3$, where its surface is described by the equality $f_3(\alpha_1,\alpha_2,\alpha_3)=0$. A tangential point satisfies \cite{m.namkung}
\begin{align}
    f_3(\alpha_1,\alpha_2,\alpha_3)=0, \ \ \frac{\partial P_s(\alpha_1,\alpha_2,\alpha_3)/\partial \alpha_j}{\partial P_s(\alpha_1,\alpha_2,\alpha_3)/\partial \alpha_k}=\frac{\partial f_3(\alpha_1,\alpha_2,\alpha_3)/\partial\alpha_j}{\partial f_3(\alpha_1,\alpha_2,\alpha_3)/\partial\alpha_k} \ \ \forall k\not=j.
\end{align}
Finding an optimal point $(\alpha_1^{\star},\alpha_2^{\star},\alpha_3^{\star})$ involves solving a 6th-order polynomial equation, which is difficult to solve analytically, except in the case of $\Phi=0$ or $\pi$ \cite{j.a.bergou}.  In particular, when the Berry phase is equal to zero, there are two types of surface solutions~\cite{j.a.bergou,m.namkung,m.namkung2}:
\begin{align}\label{surf}
    (\alpha_1,\alpha_2,\alpha_3)&\in\vec{1}-\underbrace{\left\{\left(\frac{s_2s_3}{s_1},\frac{s_1s_3}{s_2},\frac{s_1s_2}{s_3}\right)\right\}}_{\rm The \ 1st \ surface \ solution}\cup\underbrace{\left\{\left(\frac{\sqrt{q_y}s_z+\sqrt{q_z}s_y}{\sqrt{q_x}},\frac{\sqrt{q_x}s_z-\sqrt{q_z}s_x}{\sqrt{q_y}},\frac{\sqrt{q_x}s_y-\sqrt{q_y}s_x}{\sqrt{q_z}}\right):x\not=y\not=z\right\}}_{\rm The \ 2nd \ surface  \ solution},
\end{align}
with $\vec{1}=(1,1,1)$. 

\section*{Appendix B. Projective measurement}

\subsection*{B.1. Discriminating three qutrit states}
Let us consider the three quantum states on a Hilbert space spanned by an orthonormal basis $\{|l_0\rangle,|l_1\rangle,|l_2\rangle\}$:
\begin{align}\label{3p}
    |\psi_1\rangle&=\cos\xi|l_0\rangle+\sin\xi|l_1\rangle,\nonumber\\
    |\psi_2\rangle&=\cos\xi(\cos\theta|l_0\rangle+\sin\theta|l_1\rangle)+\sin\xi|l_2\rangle,\nonumber\\
    |\psi_3\rangle&=\cos\xi(\cos\varphi|l_0\rangle-\sin\varphi|l_1\rangle)+\sin\xi|l_2\rangle.
\end{align}
If the Gram matrix $\mathrm{G}$ is positive-definite, then we can construct a projective measurement on a 4-dimensional Hilbert space composed of four sub-normalized vectors
\begin{equation} \label{d}
    |D_1\rangle=|\widetilde{\psi}_1\rangle+a_1|l_3\rangle, \ \ |D_2\rangle=|\widetilde{\psi}_2\rangle+a_2|l_3\rangle, \ \ 
    |D_3\rangle=|\widetilde{\psi}_3\rangle+a_3|l_3\rangle, 
\end{equation}
where $|l_3\rangle$ is a basis vector orthogonal to $\{|l_0\rangle,|l_1\rangle,|l_2\rangle\}$ and  $a_1,a_2,a_3$ are real numbers. For orthogonality among the three vectors in Eq.~(\ref{d}), $a_1,a_2,a_3$ should satisfy
\begin{equation}
    \langle\widetilde{\psi}_1|\widetilde{\psi}_2\rangle+a_1a_2=0, \ \ \langle\widetilde{\psi}_1|\widetilde{\psi}_3\rangle+a_1a_3=0, \ \ 
    \langle\widetilde{\psi}_2|\widetilde{\psi}_3\rangle+a_2a_3=0.
\end{equation}
It is straightforward to show that the above real numbers satisfy these three equalities \cite{m.agnew}:
\begin{equation}
    a_1=\sqrt{-\frac{\langle\widetilde{\psi}_1|\widetilde{\psi}_2\rangle\langle\widetilde{\psi}_1|\widetilde{\psi}_3\rangle}{\langle\widetilde{\psi}_2|\widetilde{\psi}_3\rangle}}:=\kappa, \ \ a_2=\kappa\frac{\langle\widetilde{\psi}_2|\widetilde{\psi}_3\rangle}{\langle\widetilde{\psi}_1|\widetilde{\psi}_3\rangle}, \ \
    a_3=\kappa\frac{\langle\widetilde{\psi}_2|\widetilde{\psi}_3\rangle}{\langle\widetilde{\psi}_1|\widetilde{\psi}_2\rangle},
\end{equation}
and the normalized vectors in Eq. (\ref{d}) are 
\begin{align}\label{d123}
    |D_1\rangle=\frac{|\widetilde{\psi}_1\rangle+\kappa|l_3\rangle}{\sqrt{\langle\widetilde{\psi}_1|\widetilde{\psi}_1\rangle+\kappa^2}},  |D_2\rangle=\frac{\langle\widetilde{\psi}_1|\widetilde{\psi}_3\rangle|\widetilde{\psi}_2\rangle+\kappa\langle\widetilde{\psi}_2|\widetilde{\psi}_3\rangle|l_3\rangle}{\sqrt{\langle\widetilde{\psi}_1|\widetilde{\psi}_3\rangle^2\langle\widetilde{\psi_2}|\widetilde{\psi_2}\rangle+\kappa^2\langle\widetilde{\psi}_2|\widetilde{\psi}_3\rangle^2}}, |D_3\rangle=\frac{\langle\widetilde{\psi}_1|\widetilde{\psi}_2\rangle|\widetilde{\psi}_3\rangle+\kappa\langle\widetilde{\psi}_2|\widetilde{\psi}_3\rangle|l_3\rangle}{\sqrt{\langle\widetilde{\psi}_1|\widetilde{\psi}_2\rangle^2\langle\widetilde{\psi_3}|\widetilde{\psi_3}\rangle+\kappa^2\langle\widetilde{\psi}_2|\widetilde{\psi}_3\rangle^2}}.
\end{align}
The explicit form of these vectors is described as 
\begin{align}\label{exp_d}
 |D_1\rangle&=N_1
\Big\{
   \left(1+\cot\frac{\theta}{2}\cot\frac{\varphi}{2}\right)\sec\xi |l_0\rangle+
   \left(-\cot\frac{\theta}{2}+\cot\frac{\varphi}{2}\right)\sec\xi |l_1\rangle+
   \left(1-\cot\frac{\theta}{2}\cot\frac{\varphi}{2}\right)\csc\xi |l_2\rangle+
   \Omega(\theta,\varphi,\xi)|l_3\rangle\Big\}
 , \nonumber\\ 
  |D_2\rangle&=N_2
 \Big\{
   \Gamma(\theta,\varphi,\xi)\cot\frac{\varphi}{2}\sec\xi |l_0\rangle
   -\Gamma(\theta,\varphi,\xi)\sec\xi |l_1\rangle
   -\Gamma(\theta,\varphi,\xi)\cot\frac{\varphi}{2}\csc\xi |l_2\rangle+
   \Lambda(\theta,\varphi,\xi)\Omega(\theta,\varphi,\xi)\sec^2\xi|l_3\rangle
 \Big\}, \nonumber\\
 |D_3\rangle&=N_3
 \Big\{
   \Gamma(\theta,\varphi,\xi)\cot\frac{\varphi}{2}\sec\xi |l_0\rangle
   +\Gamma(\theta,\varphi,\xi)\sec\xi |l_1\rangle-\Gamma(\theta,\varphi,\xi)\cot\frac{\varphi}{2}\csc\xi |l_2\rangle+
   \Lambda(\theta,\varphi,\xi)\Omega(\theta,\varphi,\xi)\sec^2\xi|l_3\rangle
 \Big\},
\end{align}
with normalization constants $N_j$, and
\begin{align}
    \Gamma(\theta,\varphi,\xi)&:=\cot\frac{\varphi}{2}\sec^2\xi-\cot\frac{\theta}{2}\csc^2\xi+\cot^2\frac{\theta}{2}\cot\frac{\varphi}{2}\csc^2\xi\sec^2\xi,\nonumber\\
    \Lambda(\theta,\varphi,\xi)&:=-1+\cot\frac{\theta}{2}\cot\frac{\varphi}{2}\csc^2\xi,\nonumber\\
    \Omega(\theta,\varphi,\xi)&:=\sqrt{-\frac{\sec^2\xi\left\{\cot\frac{\varphi}{2}\left(1+\cot^2\frac{\theta}{2}\csc^2\xi\right)-\cot\frac{\theta}{2}\cot^2\xi\right\}\left\{\cot\frac{\theta}{2}\left(1+\cot^2\frac{\varphi}{2}\csc^2\xi\right)-\cot\frac{\varphi}{2}\cot^2\xi\right\}}{\Lambda(\theta,\varphi,\xi)}}.
\end{align}
The projective measurement elements $\hat{\Pi}_1$, $\hat{\Pi}_2$, and $\hat{\Pi}_3$ are constructed as $\hat{\Pi}_j=|D_j\rangle\langle D_j|$, and the element $\hat{\Pi}_?$ is constructed as $\hat{\Pi}_?=\hat{I}-\hat{\Pi}_1-\hat{\Pi}_2-\hat{\Pi}_3$ by the completeness condition. In the case of $\theta=2\pi/3$ and $\xi=\pi/3$, as considered in Figs.~3 and 4 of the main text, the measurement vectors of Eq.~(\ref{d123}) have the explicit form:
\begin{align}
    |D_1\rangle&=N_1
\left\{
   \frac{2}{\sqrt{3}}\left(\cot\frac{\varphi}{2}+\sqrt{3}\right) |l_0\rangle+
   2\left(\cot\frac{\varphi}{2}-\frac{1}{\sqrt{3}}\right) |l_1\rangle+
   \frac{2}{\sqrt{3}}\left(1-\frac{1}{\sqrt{3}}\cot\frac{\varphi}{2}\right) |l_2\rangle+
   \Omega(\varphi)|l_3\rangle\right\}
 , \nonumber\\ 
  |D_2\rangle&=N_2
 \left\{
   2\Gamma(\varphi)\cot\frac{\varphi}{2} |l_0\rangle
   -2\Gamma(\varphi) |l_1\rangle
   -\frac{2}{\sqrt{3}}\Gamma(\varphi)\cot\frac{\varphi}{2} |l_2\rangle+
   4\Lambda(\varphi)\Omega(\varphi)|l_3\rangle
 \right\}, \nonumber\\
 |D_3\rangle&=N_3
 \left\{
   2\Gamma(\varphi)\cot\frac{\varphi}{2} |l_0\rangle
   +2\Gamma(\varphi) |l_1\rangle
   -\frac{2}{\sqrt{3}}\Gamma(\varphi)\cot\frac{\varphi}{2} |l_2\rangle+
   4\Lambda(\varphi)\Omega(\varphi)|l_3\rangle
 \right\},
\end{align}
with the simplified parameters
\begin{align}
    \Gamma(\varphi)=\frac{4}{9}\left(13\cot\frac{\varphi}{2}-\sqrt{3}\right),  \Lambda(\varphi)=\frac{4\cot\frac{\varphi}{2}}{3\sqrt{3}}-1, \Omega(\varphi)=\frac{2}{3}\sqrt{\frac{9-42\sqrt{3}\cot\frac{\varphi}{2}+51\cot^2\frac{\varphi}{2}-52\sqrt{3}\cot^3\frac{\varphi}{2}}{4\sqrt{3}\cot\frac{\varphi}{2}-9}}.
\end{align}

\begin{figure*}
\centerline{\includegraphics[width=0.6\linewidth]{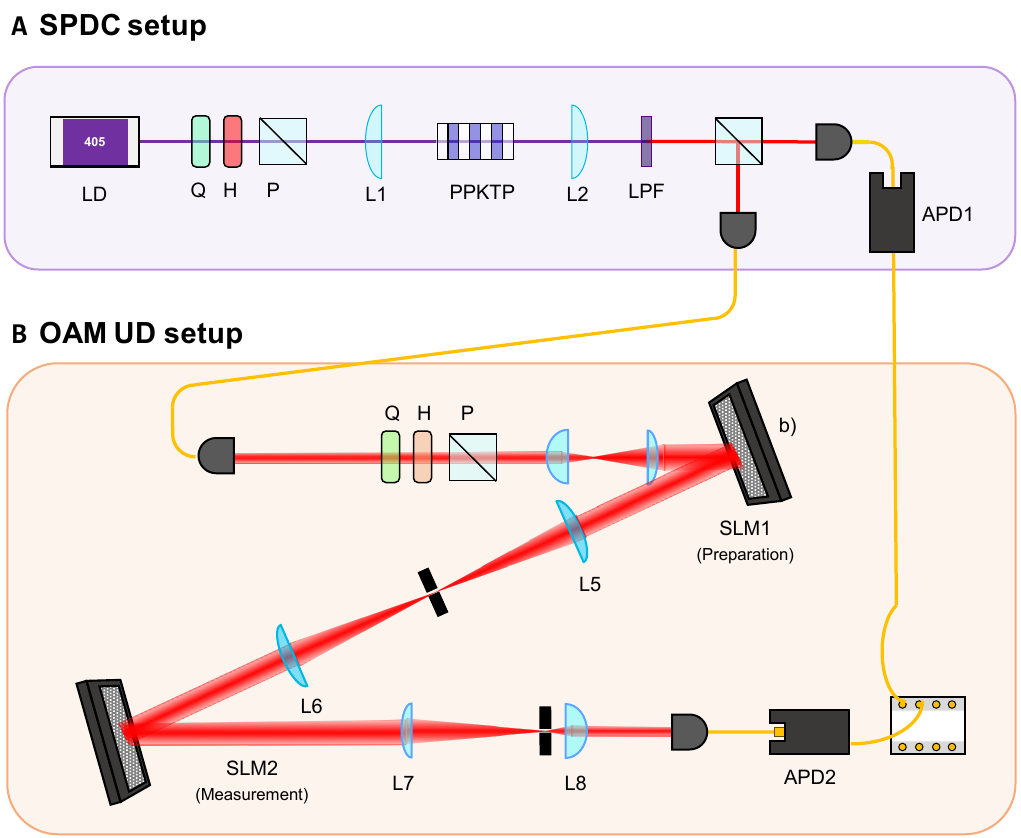}}
\caption{A detailed experimental setup for OAM USD. (a) SPDC setup for heralded single-photon generation. (b) OAM USD setup for state discrimination. (LD, Laser Diode; H, Half-wave plate; Q, Quarter-wave plate; P, Polarizing beam splitter; L, plano-convex lens; LPF, Long pass filter; SLM, Spatial light modulator; APD, Avalanche photodiode; TCSPC, Time-correlated single-photon counter).}
\centering
\label{fig6}
\end{figure*}

\begin{figure*}[t]
\centerline{\includegraphics[width=0.8\linewidth]{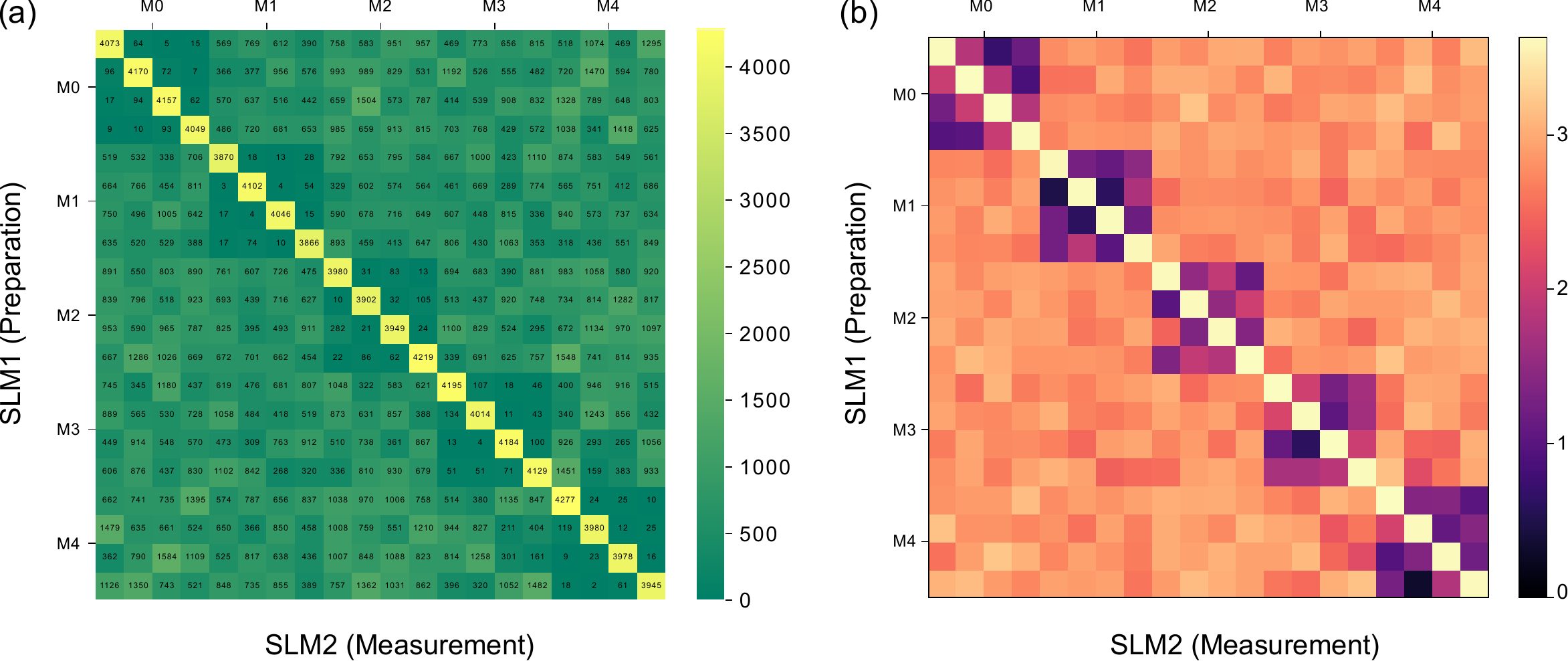}}
\caption{Experimental crosstalk matrix for 4 dimension MUB bases. Each bases in same group is orthogonal each other. (a) linear and (b) log scale.}
\centering
\label{fig7}
\end{figure*}

\subsection*{B.2. Discriminating high-dimensional qudit states}
{To discriminate $d$ quantum states without error, a high-dimensional projective measurement is composed of $d$ orthonormal vectors:
\begin{equation}\label{dj}
|D_j\rangle=\frac{1}{\sqrt{\langle\widetilde{\psi}_j|\widetilde{\psi}_j\rangle+\sum_{k=1}^{d_{\rm ext}}{a_j^{(k)2}}}}\left\{|\widetilde{\psi}_j\rangle+\sum_{j=1}^{d_{\rm ext}}a_j^{(k)}|l_{d+k-1}\rangle\right\},
\end{equation}
with real numbers $a_j^{(k)}$ such that
\begin{align}\label{eq_4p}
    \langle\widetilde{\psi}_j|\widetilde{\psi}_{j'}\rangle+\sum_{k=1}^{d_{\rm ext}}a_j^{(k)}a_{j'}^{(k)}=0 \ \ \forall j\not=j.
\end{align}
Additionally, a normalized $|D_?\rangle$ is orthogonal to these $d$ vectors in Eq.~(\ref{dj}). The optimal projective measurement is a solution of the following optimization problem:
\begin{eqnarray}
    \mathrm{maximize}&& \ \ P_s=\sum_{j=1}^{d}q_j\left|\langle D_j|\psi_j\rangle\right|^2,\nonumber\\
    \mathrm{subject \ to}&& \ \ \langle\widetilde{\psi}_j|\widetilde{\psi}_{j'}\rangle+\sum_{k=1}^{d_{\rm ext}}a_j^{(k)}a_{j'}^{(k)}=0 \ \ \forall j\not=j',
\end{eqnarray}
in which optimal parameters $a_j^{(k)}$ are evaluated using numerical approaches such as Powell method with sufficient repetitions, as the above problem is not convex~\cite{s.boyd}.}
\end{widetext}

\begin{figure*}
\centerline{\includegraphics[width=\linewidth]{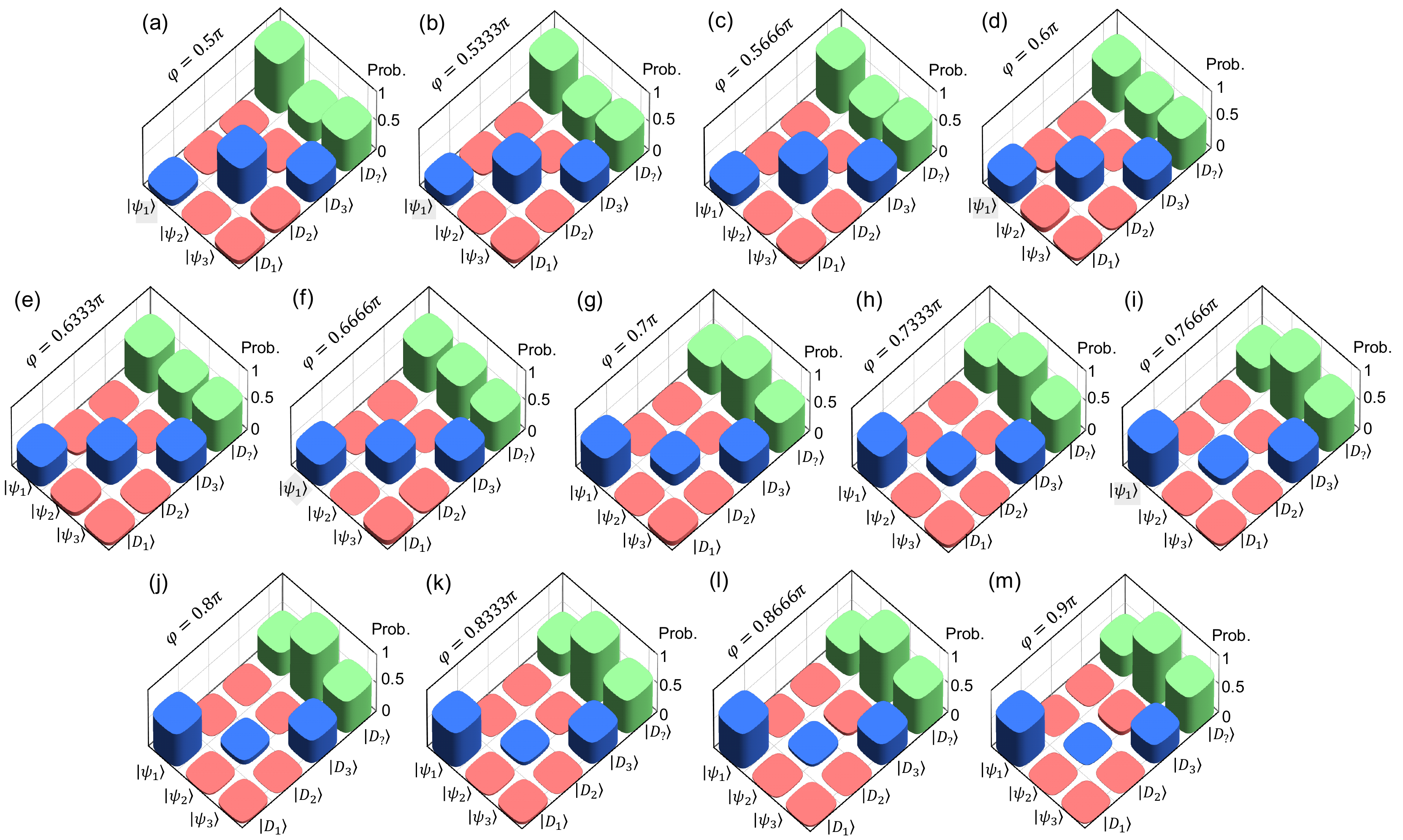}}
\caption{Experimentally obtained probability distribution for $\theta=\frac{2\pi}{3},\xi=\frac{\pi}{3}$ and each $\varphi$. (a) $\varphi=0.5\pi$, (b) $\varphi=0.5333\pi$, (c) $\varphi=0.5666\pi$, (d) $\varphi=0.6\pi$, (e) $\varphi=0.6333\pi$, (f) $\varphi=0.6666\pi$, (g) $\varphi=0.7\pi$, (h) $\varphi=0.7333\pi$, (i) $\varphi=0.7666\pi$, (j) $\varphi=0.8\pi$, (k) $\varphi=0.8333\pi$, (l) $\varphi=0.8666\pi$, and (m) $\varphi=0.9\pi$, respectively. Here, all the above nine cases consider some $\varphi\in[0.5\pi,0.9\pi]$ and $\theta=\frac{2\pi}{3}$. Probabilities to obtain a measurement outcome corresponding to $|D_j\rangle$ from a given state $|\psi_k\rangle$ are colored in blue, red, and green bars corresponding to success, error, and failure cases, respectively.}
\centering
\label{fig8}
\end{figure*}

\section*{Appendix C. Experimental implementation}

\subsection*{C.1. Details on experimental setup}

An experimental setup for USD among asymmetric qutrit states using the orbital angular momentum (OAM) state of a single-photon is illustrated in Fig. ~\ref{fig6}. A 405 nm continuous wave laser (OBIS LX FP) pumps a 10-mm-long periodically-poled potassium titanyl phosphate (PPKTP) crystal with 48 mW of power. A pair of photons, called the signal and idler photons, is then generated via a type-II spontaneous parametric down-conversion (SPDC) process. The pair of photons is separated by a polarizing beam splitter (PBS); the idler photon is sent directly to the avalanche photon detector (APD) for heralding, and the signal photon is injected into the OAM USD setup. The signal photon is horizontally polarized by a half-wave plate, quarter-wave plate, and PBS, and it is expanded to approximate a plane wave. One of the states for discrimination is prepared by a spatial light modulator (SLM, Holoeye Pluto-2.1 NIR) with the superimposed phase of Laguerre-Gaussian (LG) modes. The holographic image on SLM1 has a blazed grating, diffracting the light to separate it from the reflected light. The prepared state passes through the lens-iris-lens system, known as the 4$f$ system, where the center of the iris filters out all other diffraction orders except the desired first-order diffraction. The state is measured by SLM2 with projective measurement images. The far-field distribution of light is converted to Gaussian functions by a two-dimensional Fourier transform, and it is coupled into a single-mode fiber with Gaussian mode projection~\cite{n.bent}. Using a time-correlated single-photon counter (TCSPC) with a 2 ns coincidence window, the two photons with a different time delay can be measured in the form of a coincidence count.

\begin{figure*}
\centerline{\includegraphics[width=0.7\linewidth]{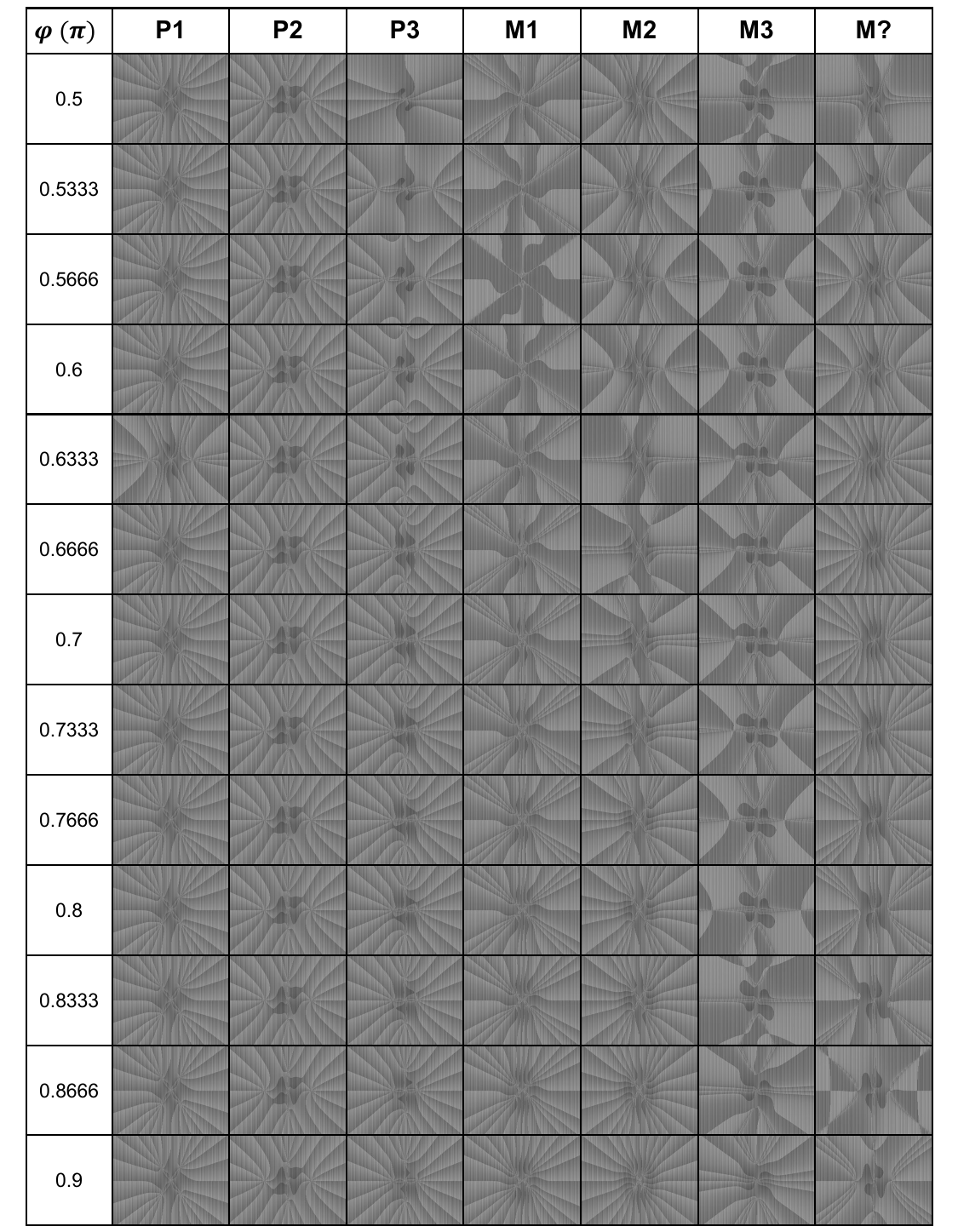}}
\caption{Phase masks for experiment of unambiguous state discrimination. Here, $\mathrm{\{P1,P2,P3\}}$ and $\mathrm{\{M1,M2,M3,M?\}}$ correspond to $\{|\psi_1\rangle$, $|\psi_2\rangle$, $|\psi_3\rangle\}$ and $\{|D_1\rangle,|D_2\rangle,|D_3\rangle,|D_?\rangle\}$, respectively.}
\centering
\label{fig9}
\end{figure*}

\subsection*{C.2. Crosstalk matrix}

To accurately prepare the states with overlap, we confirmed the orthogonality of mutually unbiased bases (MUBs), as represented in Table~S1~\cite{a.klapp}. Since the projective measurement has 4 dimensions, the MUB will consist of 20 by $d$ bases, with $(d+1)$ groups, called M0 to M4. The MUBs are orthogonal to each other within each group, meaning that the measurement result is 1 or 0. When measured in different groups, the result becomes $\frac{1}{d}$. To check orthogonality, we introduced a crosstalk matrix, which is a $d(d+1) \times d(d+1)$ matrix that verifies the preparation and measurement of LG modes~\cite{m.rambach}. Each element of the crosstalk matrix is a measurement outcome of two MUBs $| \langle \psi_m | \psi_p \rangle |^2$, where $| \psi_p \rangle$ is the prepared MUBs and $\langle \psi_m |$ is the measured MUBs. Each row and column of the crosstalk matrix correspond to the preparing and measuring MUBs on the SLM1 and SLM2. Through precise alignment, the orthogonal terms can be made as close to 0 as possible. Then visibility $V$ of each groups of MUBs shows how well the superposed LG modes are generated and measured. It is given by
\begin{equation}
	V=\frac{\sum_{i=1}^{d}N_{ii}}{\sum_{i,j=1}^{d}N_{ij}},
\end{equation}
where $\sum_{i=1}^{d}N_{ii}$ is the sum of the coincidence counts of the diagonal elements of each group of MUBs, and  $\sum_{i,j=1}^{d}N_{ij}$  is the sum of all coincidences of each group for prepared MUBs $i$ and measured MUBs~\cite{m.rambach}. We achieved visibilities for each group, $V \in \{96.8, 98.4, 95.4, 96.2, 97.9\}$ $(\%)$, as shown in Fig.~\ref{fig7}. We confirmed the small crosstalk counts between orthogonal superposed LG modes, and these nonzero counts can cause a slight experimental error probability when discriminating the asymmetric states.

\subsection*{C.3. Measurement probabilities}
Here, we present experimental data for various cases of the USD in Fig.~\ref{fig8}, which are included in Fig. 4 of the main text. It is observed that the error probabilities (red bars) are close to zero. Thus, the results shown in Fig.~\ref{fig8} confirm the successful demonstration of USD.

\subsection*{C.4. Phase masks for experiment}
Here, we provide the phase masks for both state preparation and measurement through experiment in Fig.~\ref{fig9}. These holographic images can be generated by manipulating helical phase of LG modes~\cite{c.rosales}, diffraction grating~\cite{w.shao}, and intensity masking~\cite{e.bolduc}. We note that the grating direction for preparation and measurement are opposite to each other, thereby canceling out the diffraction effect of the momentum.


\begin{thebibliography}{99}
\bibitem{note}
We note that the symmetry according to our definition is generally different from the geometrical symmetry \cite{a.chefles2}. According to our definition, qudit states $|\psi_1\rangle,\cdots,|\psi_d\rangle$ are symmetric when all overlaps $\langle\psi_j|\psi_{j'}\rangle$ are equal to each other. Notably, this is not guaranteed by a unitary operator $\hat{V}$ that satisfies $|\psi_{j+1 \mathrm{ \ mod \ } d}\rangle = \hat{V}|\psi_j\rangle$. This can be verified from a counterexample of qudit states composing phase-shift-keying signal \cite{f.e.becerra}.
\end{thebibliography}
\end{document}